# MetaZipf. (Re)producing knowledge about city size distributions.

Clémentine Cottineau
Centre for Advanced Spatial Analysis - University College London
90 Tottenham Court Road, London W1T 4TJ, UK c.cottineau@ucl.ac.uk

**Abstract.**
Zipf's law for cities is probably the most famous regularity in social sciences. So much that, a hundred years of publication later, its status is not clear: is it a law of social organisation? Is it an instrument of description of city size distributions? Is it an element of validation of geographical objects (cities and systems of cities in particular)? Empirical estimations of the rank-size parameters are very numerous and contradict each other. In this study, we present the results of a reproducible meta-analysis of the largest pool of papers regarding this issue, obtained from the collection of data made open and the construction of an online interactive application which allows the reader to explore this literature. We find that a large part of the variations observed in the measure of Zipf's coefficient is unnecessary as it comes from the choice of different technical specifications in the way cities are defined, whereas some of the current theories to explain the remaining share of variations are challenged by our results.

**Keywords.**
Zipf, Cities, Meta Analysis, Deviations, Open Data, Urban Theory

> *"It is certainly not clear just what are the logical links between the scheme proposed by Zipf to explain rank-size regularity and observed rank-size regularities. Thus, it would not be proper to credit Zipf with an articulated empirical and theoretical analysis of the rank-size problem."*
> Berry and Garrison, 1958, p.85

## 1. INTRODUCTION

Since the discovery of a mathematical regularity in the distribution of city populations by Auerbach [1913] and the claim of its exact form (a power law) and scaling parameter (-1) by Zipf [1949], there has been a continuous flow of empirical research confirming or challenging this result. The rank-size distribution of city populations, often summarized as "Zipf's Law", has focused the attention of a variety of disciplines such as geography, economics, linguistics, archaeology, mathematics, physics, etc. in a large diversity of spatio-temporal urban contexts and under many different sets of assumptions, regarding the measure of size, the definition of cities or the factors generating the regularity. This field of research is still on-going, as attested by the fact that more than half of the studies included in this review, 42 out of 82, were published between 2006 and 2016.

An important strand of research related to the empirics of Zipf's law aims at finding the right way to estimate the relation between ranks and sizes. This involves challenging the power law as the mathematical form of the relation (in favour of a log-normal law for example [Eeckout, 2004]), the OLS as the right method of estimation (in favour of the maximum likelihood Hill estimator [Soo, 2005]), as well as the expression of rank (in favour of rank - 1/2 [Gabaix and Ibragimov, 2011] for example). Such heterogeneity produces semi-comparable results and does not explain the variation of the parameter value for a given method depending on other factors [Rosen, Resnick, 1980; Soo, 2005; Nitsch, 2005].

**Table 1. Variation of Zipf's law in the Former Soviet Union with specifications**

| Cities as... | Min Pop | Year | α | Sd | R²(%) | N | Graph |
|---|---|---|---|---|---|---|---|
| **Local Units** | 10,000 | 2010 | 1.046 | 0.003 | 98.4 | 2064 | 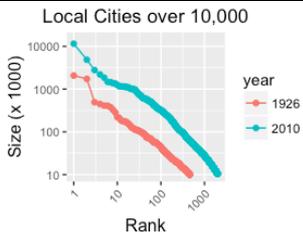 |
| | | 1926 | 0.856 | 0.003 | 99.4 | 461 | |
| | 100,000 | 2010 | 0.796 | 0.008 | 96.9 | 297 | 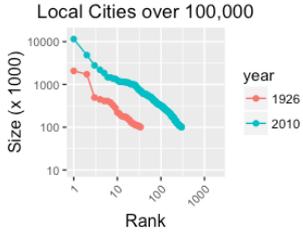 |
| | | 1926 | 0.856 | 0.029 | 96.5 | 34 | |
| **Built-up Areas** | 10,000 | 2010 | 1.104 | 0.003 | 98.7 | 1925 | 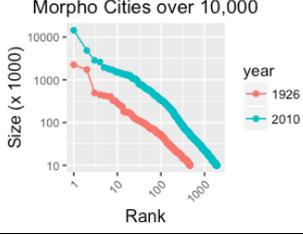 |
| | | 1926 | 0.878 | 0.004 | 99.0 | 470 | |
| | 100,000 | 2010 | 0.826 | 0.009 | 96.5 | 278 | 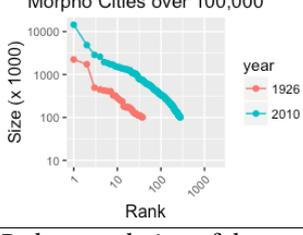 |
| | | 1926 | 0.826 | 0.026 | 96.6 | 38 | |

Source: [Cottineau, 2014]. Estimation of α from: $\log(P_i) = \alpha \log(i) + \beta$, with $P_i$ the population of the city of rank i.

For example, we estimated Zipf's law for the cities of the Former Soviet Union with a standard OLS regression on full ranks, using two different city definitions, two population cutoffs and for two different years (table 1):

$$\log(P_i) = \beta - \alpha \log(i) \qquad (1)$$

with $P_i$ the population of the city of rank i.

When cities are defined as administrative units with an urban status ('Local Units') over 10,000 residents, the estimation of the parameter α in equation (1) indicates an increase in the level of population unevenness between 1926 (α = 0.856) and 2010 (α = 1.046). However, if we consider only cities over 100,000 residents at each time period, the estimation is reduced respectively to 34 and 297 cities, and gives the opposite result: the parameter α in this case shows a decrease in value, from 0.856 to 0.796. This means that the population cutoff did not affect the measurement of the rank-size in 1926 (α = 0.856 at both dates), but dramatically affected the measurement in 2010. Additionally, when considering cities as built-up areas in the post-Soviet space [Cottineau, 2014], one finds a similar result with the population cutoff of 10,000 but a stable value over time with the cutoff of 100,000 (where both α = 0.826).

This example confirms the observation that the estimation of Zipf's law varies largely with respect to the system's definition. Therefore, how to assess the validity of theories based on this law? What do deviations tell us about the process of urbanisation in different urban contexts? Two approaches have been taken to explain the variation in Zipf's coefficient estimations. The first approach consists in gathering data on cities for a diversity of countries, dates and specifications [Rosen, Resnick, 1980; Parr, 1985; Moriconi-Ebrard, 1993; Soo, 2005]. The second approach consists in conducting a meta-analysis of the literature [Nitsch, 2005]. In this paper, we present an analysis of the second sort, which relies on articles of the first sort as it tries to review systematically as many systematic studies as possible.

Our analysis adds to the existing literature in that we provide a more extensive review of the literature, potentially enhanced by collective participation. Indeed, V. Nitsch [2005] used 512 estimations of α from 29 studies. We propose an analysis of 1702 estimations from 81 studies (cf. Appendix A). Furthermore, the meta-analysis itself is transparent and adaptable because it is supported by a dedicated online tool, MetaZipf[1], which allows to reproduce, to explore and to visualise the results (cf. Appendix C). Finally, the data gathered for this study are provided under an open-licence[2] for a cumulative building on knowledge around Zipf's law.

## 2. A HUNDRED YEARS OF PUBLICATION ON ZIPF'S LAW

The data for a meta-analysis consists in comparable studies published by a diversity of authors. We found 81 such studies where a value of α was estimated, either in the Lotka form of equation (1), either in the Pareto equation (where the rank i is regressed as a function of population $P_i$), using ranks or ranks - 1/2, without any other variables included in the regression. To make Pareto estimates comparable with Lotka coefficients here, we calculate their reciprocal and express all results in the Lotka form. As pointed out by V. Nitsch in his own review of Zipf's estimates, the choice of studies included in the meta-analysis has a strong impact on the results obtained. Our goal is to include as many studies as possible, using manual search in the present version, and potentially crowsourcing through the online application or the network of citations from Google Scholar based on articles with similar keywords[3] in a future version. We present an overview of the literature collected before producing a summary of the distribution of estimates.

2.1. Literature overview

The empirical study of Zipf's law for cities is a classical subject of regional and urban science. Indeed, among the 81 studies included in the review, 8 were published in *Urban Studies*, 6 in the *Journal of Regional Science*, 5 in the *Journal of Urban Economics* and 5 in *Regional Science and Urban Economics*. Most of the studies (42 out of 82) were published during the last ten years, although there seems to have been a fashion as well in the 1980s, with 13 studies published between 1980 and 1995 (figure 1A).

---

[1] https://clementinegeo.shinyapps.io/MetaZipf
[2] https://github.com/ClementineCttn/MetaZipf
[3] https://github.com/JusteRaimbault/MetaZipf

**Figure 1. Distribution of studies and estimates**

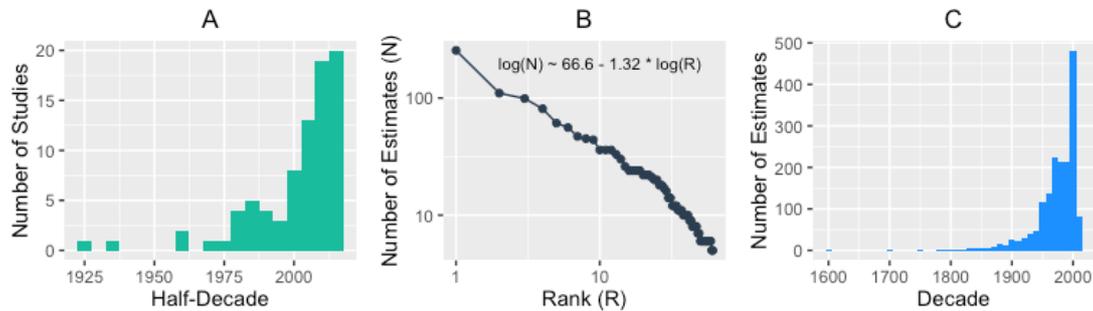

A: Histogram of studies by half-decade of publication. B: Rank-size plot of the number of estimate by study reporting more than 5 estimates. C: Histogram of estimates by decade of the original urban data.

The distribution of the number of estimates by study is very skewed itself, with few authors providing a very large number of estimates and the majority of them computing half a dozen values (figure 1B). The studies with the most estimates published are comparative studies with estimates for many countries in the world (Moriconi-Ebrard [1993]: 255 estimates, Soo [2005]: 99 estimates; Parr [1985]: 61; Rosen and Resnick [1980]: 56). Another type of studies with a large number of estimates corresponds to sensitivity analyses on a single territory, with variations of dates and cutoff values (Gonzalez-Val [2010]: 110 estimates in the United States, Xu & Zhu [2009]: 45 estimates in China). Those studies have the inconvenient of being over-represented in the meta-analysis when one estimate counts as one observation. However, they have the advantage of providing comparable cases to better isolate the effect of unique factors (minimum population, year, etc.), everything else being equal. We therefore weight estimates equally, irrespective of the study in which they where published. The distribution of the number of estimates measured over time reveals the increased availability of urban data, first in 1950, where the series published by the UN start, then in the 1970s (when most national censuses of developed countries and countries from the Commonwealth become accessible) and the 2000s, with almost 500 estimates for this decade, including those relative to developing countries (figure 1C).

2.2. Distribution of estimates

Overall, we find the distribution of $\alpha$ values estimated in the literature to be centred on 1 and relatively symmetrical (figure 2A). More than 70 estimates corroborate Zipf's prediction very precisely, with an estimation of $\alpha$ comprised in the interval ]0.99 ; 1.01[ (cf. Appendix B). However, this distribution also reveals a large dispersion around the value 1: the standard deviation of the distribution is 0.269, for a mean estimate of 1.031 (table 2). Only 628 out of 1702 estimates expressed in the Lotka form are comprised in the interval [0.9; 1.1], and 322 in the interval [0.95 ; 1.05]. The minimum value of 0.269 was estimated for 5 cities defined as local units over 100,000 residents in 1951, in the Indian region of Andhra Pradesh, by Kumar and Subbarayan [2014]. The maximum value of 3.846 was estimated on 142 cities defined as built-up areas (morphoCities) over 3,000 residents, in 1950 in China, by Luckstead and Devadoss [2014].

The way cities are defined thus seems to affect the value obtained: the distribution of cities defined as local units is clearly centred on lower values than morphological cities (figures 2B and 2C), a feature already noticed by Auerbach [1913], Rosen and Resnick [1980], Soo [2005] and Nitsch [2005]. For metropolitan areas (figure 2D), the picture is less clear-cup, with two modes around 1 and a lower number of estimates.

**Figure 2. Distribution of Estimates**

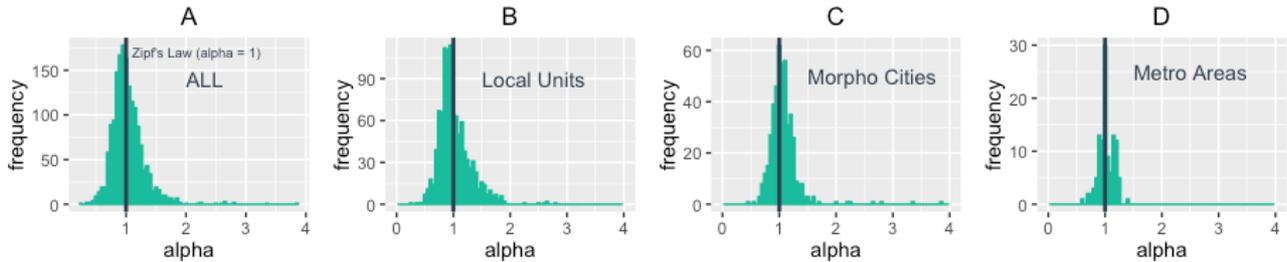

A: Distribution of all estimates expressed in Lotka form. B: Subset of the distribution for cities defined as local units. C: Subset of the distribution for cities defined as morphological cities (built-up areas). D: Subset of the distribution for cities defined as metropolitan areas.

The distribution of the values of estimates with time (table 2), shows a continuous increase in city size unevenness since the 17$^{th}$ century. On average, the coefficients measured before 1900 are much lower than 1 (0.833 based on 71 estimates from 14 studies). Using urban data from the first half of the 20$^{th}$ century, authors report an average value of 1.014, very close to the value 1 advocated by Zipf as a law in 1949. Since then, the average value of α has increased (to 1.05 for studies of cities from 1990 to 2015), indicating an accentuation of the level of inequality of city sizes over time.

**Table 2. Average coefficients of Zipf's law by category of estimations**

|  | Estimates | Studies | Mean α | sd α | Median Number of cities |
|---|---|---|---|---|---|
| all | 1702 | 81 | **1.031** | 0.269 | 122 |
| ... by City Definition* | | | | | |
| Local Unit | 1039 | 47 | **1.022** | 0.290 | 111 |
| MorphoCity | 460 | 22 | **1.070** | 0.280 | 126 |
| MetroArea | 114 | 18 | **1.019** | 0.137 | 161 |
| Other | 89 | 8 | **0.942** | 0.279 | 291 |
| ... by Period | | | | | |
| 1600-1900 | 71 | 14 | **0.833** | 0.198 | 116 |
| 1900-1950 | 165 | 26 | **1.014** | 0.213 | 157 |
| 1950-1990 | 697 | 54 | **1.032** | 0.279 | 99 |
| 1990-2015 | 769 | 59 | **1.051** | 0.293 | 142 |
| ... by Continent | | | | | |
| Europe | 524 | 36 | **0.955** | 0.211 | 116 |
| Asia | 607 | 29 | **1.029** | 0.336 | 128 |
| Africa | 100 | 14 | **1.099** | 0.184 | 62 |
| America | 451 | 34 | **1.106** | 0.259 | 210 |
| Oceania | 12 | 4 | **1.269** | 0.322 | 47 |
| WORLD | 8 | 2 | **0.737** | 0.139 | 100 |

|  | ... by Minimum Population |  |  |  |  |
|---|---|---|---|---|---|
| [0-10,000[ | 381 | 26 | **1.027** | 0.297 | 164 |
| [10,000-100,000[ | 670 | 51 | **1.040** | 0.260 | 146 |
| [100,000; +∞[ | 153 | 29 | **0.882** | 0.199 | 79 |
| Unknown | 498 | 22 | **1.067** | 0.301 | 100 |

\* The city definition categories were reconstructed and generalised from the description of urban data. **Local units** refer to administrative units refered to as urban, such as municipalities. **MorphoCities** are cities defined as the aggregation of local units using morphological criteria, such as the built-up area. **MetroArea** refer to the aggregation of local units using a functional criterion, such as the commuting flows. cf. section 3.2.1 for further details.

The number of estimates by continent reflects their difference of population levels as well as the availability of quality urban data of the different areas (table 2). Therefore, 607 estimates from 29 studies relate to Asia, 524 to Europe and 451 to America whereas only 100 estimates refer to Africa and 12 to Oceania. 2 studies have produced 8 estimates using world cities. The average value of α in each of the 5 continents tends to reflect the age of urbanisation, with the continents urbanised first (Europe, Asia) exhibiting lower levels of size inequality (lower α) compared to recently urbanised continents (America and Oceania). This observation is rather confirmed at the national scale (figure 3A): higher values on average are reported for "new" countries (Canada, USA, Autralia, South Africa) compared to the "old world" of ancient urbanisation (Italy, China, India, Egypt, etc.). However, there is a strong uncertainty for countries such as China, Australia, India or Spain where the most contradictory estimations have been published. In these four countries, the standard deviation of α exceeds 0.3, a high value in comparison to the expected value of 1 for α.

**Figure 3. Spatial summary of estimates**

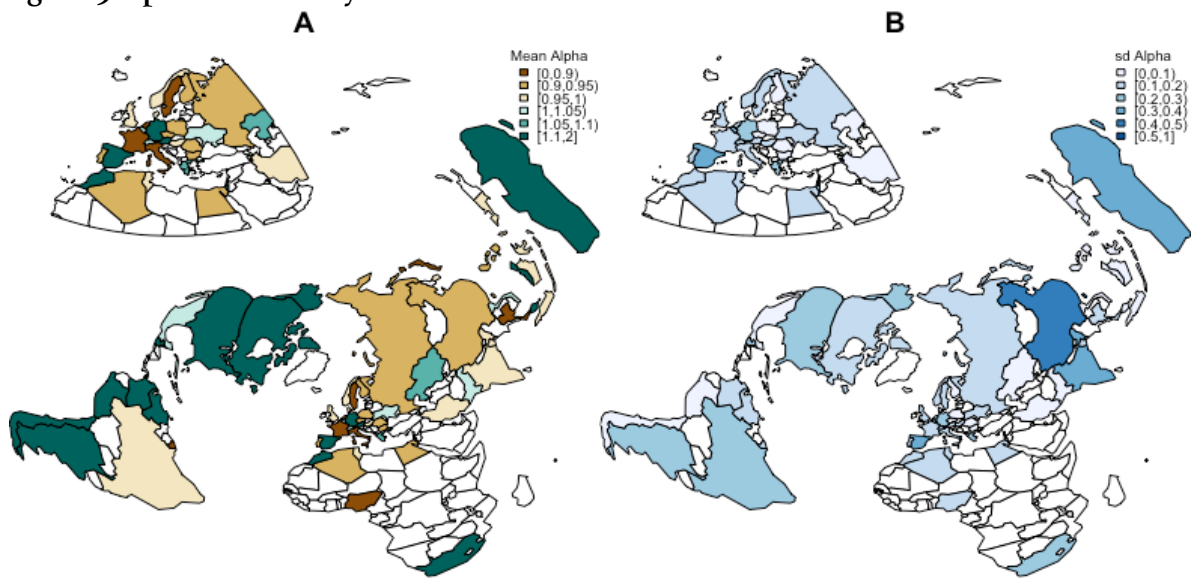

Source: Map Background from ESPON Database. A: Mean value of α by country with more that 5 estimates. B: Standard deviation of α by country with more that 5 estimates.

Finally, we find that choices in the selection of urban data can potentially affect the estimated value of α. For example, studies of the upper end of the urban hierarchy (with population *minima* of 100,000 residents or higher) exhibit a greater evenness of city sizes (α = 0.882) compared to estimates using lower-level cities (α = 1.04 for population cutoffs of 10,000 to 100,000 and α = 1.03 for cutoffs lesser than 10,000 residents, i.e. including very small cities into the analysis). However, a point worth mentioning is the poor quality of data for this variable, as authors tend to report this information less frequently than other specifications (22 out of 81 studies fail to provided the information).

## 3. WHAT MAKES EMPIRICAL RESULTS DIVERGE?

> *"Almost no data set corresponds exactly to the rank-size rule, so interpretations are based on how the data set diverges from the expected results."*
>
> Savage, 1997, p.233

It seems clear already that empirical results diverge from the parameter values predicted by G. K. Zipf. Rather than a dismissal of the theory as irrelevant, the explanation of empirical deviations from the (rank-size) rule can provide a more complete and complex understanding of the "differential forces of urbanisation" [Berry and Garrison, 1958, p.91]. Indeed, several researchers have provided explanations as to why certain deviations are observed. We review these theories before turning to the description of factors used to test them in a statistical meta-analysis.

3.1. Why divergence is expected: a review of potential explanations

Four groups of explanations can be identified from the literature, two of which relate to urbanisation processes, and two of which relate to the technical specifications of the regression used to estimate $\alpha$.

*3.1.1. Transport costs, speed and the distribution of city size in space*
Some authors have explained differences in city size unevenness by differences in transportation features at the time when city sizes are recorded. Fujita et al. [1999] argue that lower transport costs reduce the necessity for a large number of cities, since access to the agricultural products can be extended further for the same price. Moriconi-Ebrard [1993] and Pumain [1997] resort to differences in transportation speed at the time of urbanisation to explain the differences in urban hierarchies. They argue that in worlds of ancient urbanisation, the connexion between neighbouring cities were achieved at human and horse speed, thus creating a large network of small cities close to one another. By contrast, in territories urbanised through railways and motorways, there has not been the need of such a tight network and the population has clustered more unevenly in distant cities of larger size on average. This group of hypotheses is to be related to the original explanation provided by Zipf as to why the rank-size rule itself emerges: Zipf viewed the regularity as the balance of an optimisation process of location to minimize the cost of extraction and transfer of raw materials to the cities (the "force of diversification") as well as the transportation costs of processed goods to the end-consumers (the "force of unification").

*3.1.2. Concentration, integration and the size of systems*
A second group of hypotheses relates to the spatial concentration of the urban population and its integration within systems of cities. For example, Jefferson [1939], Morrill [1970], and Rosen and Resnick [1980] highlit the increased primacy of small countries and consequently their increased tendency to exhibit an uneven rank-size distribution of cities. The increased primacy in small countries would be due to the concentration of power in the capital city, which in those cases is not balanced by a sufficient set of secondary cities. A dynamic deviation from Zipf's law has also been observed, with the level of inequality of city size increasing with time [Pumain, 1997; Nitsch, 2005]. It is explained by a slight hierarchical advantage towards large cities in terms of population growth, as large cities are better equipped economically to generate and attract innovations. Finally, Harris [1970], Johnson [1980] and Rosen and Resnisk [1980] pointed to the level of integration of systems of cities to explain deviations from Zipf's law: loosely integrated systems would deviate from the value 1, generally being more even than expected, because small to medium cities would be duplicated and the primate city would not be "large enough" for the whole territory.

*3.1.3. Cities and suburbs, a definitional factor*
As a technical factor of deviation from Zipf's law, the definition of cities seems to be the most widespread explanation [Auerbach, 1913; Rosen & Resnick, 1980; Nitsch, 2005], although empirical evidence are mixed if not contradictory [Soo, 2005]. The basic argument goes as follows: if one delineates cities as built-up areas or more so as metropolitan areas, it is expected that their size distribution would be more uneven than if one delineates cities as administrative local units, because the different definitions tend to match for the smaller cities, whereas larger cities have larger suburbs and thus larger discrepancies between different city definitions. Therefore, one would record a larger population for them although their rank would remain similar.

*3.1.4. Sample size and sample cutoffs*
Finally, the size of the sets of cities used for the estimation, which is partly determined by the population cutoff chosen (the minimum population of cities), should explain some technical deviations from the rank-size rule. Indeed, small sets of cities do not represent the complete distribution of cities [Rosen and Resnick, 1980] or produce inconsistent sets [Cristelli et al., 2012] with inconsistent coefficients. On the other hand, because rank-size distributions can be slightly convex and/or concave in log-log plots [Moriconi-Ebrard, 1993; Savage, 1997; Soo, 2005], the choice of the cutoff affects the level of unevenness measured. For a convex curve, very low and very high population cutoffs will give high levels of uneveness, whereas medium cutoffs will generate low levels of unevenness measured on the same system of cities. These hypotheses have to be tested, confronted and completed, using the literature available, along with a harmonized description of regression specifications.

3.2. Results of the largest meta-analysis of Zipf's law for cities

This section details the data gathered from analysing 81 studies of Zipf's law including empirical estimations, as well as the methods and results of the meta-analysis performed.

*3.2.1. Data*
For each estimation found in a given study, we collected as many values as possible of each for the following variables:

ALPHA: The **value of** α estimated. The variation of this value is what we want to explain; therefore ALPHA is the dependent variable.

REGRESSION_FORM: The **form of the regression**, which can be the Pareto form $log(i) = α' log(P_i) + β'$ or the Lotka form $log(P_i) = α log(i) + β$. This information is used to choose one form (Lotka in this paper) and convert all the other estimates (Pareto in this paper) into their reciprocate, thus making all estimates comparable.

DATE: The **date of estimation**. This date refers to the date at which the population of cities was measured. We transform this variable into DATE = (DATE - 1950) so that the distribution is centred on 1950. We can then study the effect of the time factor relatively to 1950 (corresponding to the time of Zipf's statement). This information is used to test the hypothesis according to which systems of cities increase their level of hierarchy over time, everything else being equal (cf. section 3.1.2).

URBAN_DEF: The **definition chosen to delineate cities** in the data used in the study. This original variable is quite problematic because the authors of the studies do not provide the same level of detail in their descriptions of data. Moreover, given the diversity of countries and time periods, the original urban units ("communes", "SMA", "Boroughs", "FUA", "municipios", etc.) are hardly comparable. Therefore we create a secondary variable called URBAN_SCALE which indicates more generally how the cities are defined: as local administrative units (LocalUnits); as morphological aggregations of local units into built-up areas (MorphoCities); as functional aggregations of local units into metropolitan areas (MetroAreas) or as a mix of these or else (VariaMixed). This information is used to test the hypothesis according to which systems of composite cities have a higher level of hierarchy compared to their local counterparts, everything else being equal (cf. section 3.1.3).

TERRITORY: The **territory under observation**. This refers to the national, regional or continental territory within which urban data were collected, for example Spain, Andhra Pradesh, the OECD or any of the 164 territories we encountered in this meta-analysis. This information itself is interesting, but not usable as such. We want to describe territories with common features rather than use them as fixed effects in a regression model. Therefore, we create three secondary variables from this original one. The TERRITORY_TYPE describes the scale of administrative integration, which we classified as "regional", "national" or "macro regional" and is used to test the hypothesis according to which integrated (typically national) systems of cities are more prone to follow Zipf's law compared to other systems (cf. section 3.1.2). The variable TOTAL_POPULATION refers to the total number of residents in the territory under observation. This information is mostly available for national States between 1950 and 2015 (using the UN yearly population estimates[4]) and we discretised it into three categories: small territories under 10 million people, medium territories between 10 and 100 million, and large territories over 100 million. This new variable is used to test the hypothesis according to which small systems of cities are more uneven than large ones (cf. section 3.1.2). Finally,

---

[4] http://esa.un.org/unpd/wpp/DVD/Files/1_Indicators%20(Standard)/EXCEL_FILES/1_Population/WPP2015_POP_F01_1_TOTAL_POPULATION_BOTH_SEXES.XLS

URBAN_AGE is a variable which indicates if the territory is located in an area where urbanisation is ancient or not. 'OLD' continents refer to zones of early urbanisation, in Europe, South-East Asia and the Middle East. America, Oceania, Africa and central Asia are considered 'NEW' in that respect. This information is used to test the hypothesis according to which systems of cities in territories of ancient urbanisation are less uneven than systems of cities from the new world (cf. section 3.1.1).

N: The **number of cities** used in the regression. This information was discretised into three categories: small samples under 30 cities, medium samples between 30 and 300 cities, and large samples over 30 cities. This new variable is used to test the hypothesis according to which regressions performed on complete systems of cities are more prone to follow Zipf's law compared to regressions performed on a subset of cities only (cf. section 3.1.4).

POPULATION_CUTOFF: The **minimum population of the cities** used in the regression. This information was discretised into three categories: low cutoffs under 10,000 residents as a minimum population, medium cutoffs between 10 and 100,000, and high cutoffs over 100,000. This new variable is used to test the hypothesis according to which regressions performed on complete systems of cities are more prone to follow Zipf's law compared to regressions performed on a subset of cities only (cf. section 3.1.4).

*3.2.2. Models, Results and hypotheses testing*
We implemented three models of multiple regressions to investigate the statistical relationships between the independent variables and the value of α reported in he study.

Urbanisation Model
$$ALPHA_k = b_0 + b_1 * URBAN\_AGE_k + b_2 * DATE_k + b_3 * TOTAL\_POPULATION_k + b_4 * TERRITORY\_TYPE_k + e_k$$

Technical Model
$$ALPHA_k = b_0 + b_5 * URBAN\_SCALE_k + b_6 * POPULATION\_CUTOFF_k + b_7 * N_k + e_k$$

Complete Model
$$ALPHA_k = b_0 + b_1 * URBAN\_AGE_k + b_2 * DATE_k + b_3 * TOTAL\_POPULATION_k + b_4 * TERRITORY\_TYPE_k + b_5 * URBAN\_SCALE_k + b_6 * POPULATION\_CUTOFF_k + b_7 * N_k + e_k$$

with $b_0$ the intercept of the regression, $b_{i,\,i>0}$ the coefficients associates with the variables of interest and $e_k$ the term of error (or residual of an estimation k reported from the literature).

The Urbanisation Model includes all the variables which translate hypotheses about the urbanisation processes, i.e. the urbanisation age (ancient or new world), the date of observation (relative to 1950), the territory population and its type (region, State or macro-region). The Technical Model comprises only the variables which translate "technical" hypotheses, i.e. the effects of technical specification choices on the variation of α. This model thus includes the city definition (local, morphological, metropolitan or mixed) and cutoff, as well as the number of cities. The Complete Model includes all the variables. All models are estimated with Ordinary Least Squares. Their analysis allows to test the four groups of hypotheses we identified as well as to **disentangle the topical reasons**, i.e. explanations of why some systems of cities are more unequal than others with respect to size, **from the technical reasons** of a variation of α, as the latter could be be corrected by researchers to produce comparable estimates. All results are reported in table 3.

First, we note that the variables translating topical explanations alone fail to model the variations in α observed in the literature. Indeed, they account only for 4.5% of the variance of the 1422 values of α reported with these specifications. The technical model performs still poorly but better ($R^2$ = 11.4%), whereas about a quarter of the information (23.1%) is modelled by the combination of all variables in the complete model on 559 estimates. This summary of the models performance indicates that technical specifications play a large and unnecessary role in the mixed evidence regarding the estimation of Zipf's law. On the positive side, it means that, by using similar city definitions and population cutoffs, it would be easier to obtain comparable results and to extend the study of spatio-temporal differences in urban unevenness, the remaining 90% of the variations of α!

Although the average value of α for all 1702 estimates from the literature is 1.031, the average value for the reference cases in all three models is much lower. It is 0.945 for studies estimating α on cities of large nation States in areas of ancient urbanisation in 1950 (i.e. in the reference case for the Urbanisation Model). It is 0.839 for studies estimating α over more than 300 cities defined as local units of at least 100,000 residents (reference case for the Technical Model). It is 0.983 for studies estimating α over more than 300 cities defined as local units of at least 100,000 residents, in large nation States in areas of ancient urbanisation in 1950 (reference case for the Complete Model).

With reference to these categories, we confirm three theoretical hypotheses about the systematic deviations of Zipf's coefficients. Firstly, urban systems of recent urbanisation appear significantly and consistently more uneven, adding around 0.1 to the value of α in the Urbanisation and Complete models, everything else being equal. Secondly, in accordance with theoretical expectations, estimating Zipf's law on urban aggregations rather than on local units increases the level of size inequality measured. Within the Technical Model, this result holds equally for both morphological cities and metropolitan areas, with an average α greater by 0.14 on average. Thirdly, we find that the population cutoff used to select cities is of high importance in the Technical and Complete Models. Indeed, on average and everything else being equal, regressing Zipf's law on the complete set of cities results correlates with finding a coefficient greater by 0.1 to 0.2 compared to regressing the same law on the largest cities only (> 100,000 residents). This result confirms the existence of convex rank-size curves and it points to the lack of representativity of small sets of cities with respect to the complete distribution.

Table 3. Meta Analysis of Zipf's law

| Variable | Urbanisation Model | Technical Model | Complete Model |
|---|---|---|---|
| Intercept $b_0$ | 0.945 *** (0.023) | 0.839 *** (0.032) | 0.983 *** (0.048) |
| **Urban Age** | | | |
| Old | | ref. | |
| Recent | + 0.127 *** (0.017) | | + 0.085 *** (0.026) |
| **Date - 1950** | n.s. | | -0.004 *** (0.001) |
| **Total Population** | | | |
| Small (<10 million) | + 0.047 * (0.025) | | n.s. |
| Medium (10-100 million) | + 0.040 ** (0.019) | | n.s. |
| Large (>100 million) | | ref. | |
| **Territory Type** | | | |
| Region | + 0.085 *** (0.021) | | - 0.153 *** (0.040) |
| Nation State | | ref. | |
| Macro-region | n.s. | | - 0.276 *** (0.102) |
| **Urban Scale** | | | |
| Local Unit | | ref. | |
| MorphoCity | | + 0.143 *** (0.025) | + 0.183 *** (0.031) |
| MetroArea | | + 0.140 *** (0.025) | n.s. |
| VariaMixed | | n.s. | n.s. |
| **Population Cutoff** | | | |
| Low (<10,000) | | + 0.173 *** (0.029) | + 0.228 *** (0.038) |
| Medium (10-100,000) | | + 0.080 ** (0.033) | + 0.138 *** (0.039) |
| High (>100,000) | | ref. | |
| **Number of Cities** | | | |
| Small (<30) | | + 0.060 * (0.035) | n.s. |
| Medium (30-300) | | n.s. | n.s. |
| Large (>300) | | ref. | |
| N Observations | 1422 | 718 | 559 |
| $R^2$ (%) | 4.5 | 11.4 | 23.1 |

*ref.* = reference category for this variable. *n.s.* = Non-Significant coefficient. * p-value < 0.1, ** p-value < 0.05, *** p-value < 0.01. In parentheses: standard deviation of the coefficient estimated.
N.B. Most variables have been discretised into groups to ease interpretation. The bounds to these categories can be varied on the online tool MetaZipf https://clementinegeo.shinyapps.io/MetaZipf/

The hypothesis of an increase of unevenness over time, although apparently confirmed in table 2, is rejected here. The effect of the date is non-significant in the Urbanisation Model, and plays the opposite way in the Complete Model, with systems of cities getting slightly more even on average with time. The hypothesis relative to the size of the surrounding territory is rejected as we find similar average values of α in large, small and medium countries. This result is not the most surprising though because the hypothesis relies mostly on the observation of the capital city, whereas Zipf's law is estimated on the complete set of cities, each being given a similar weight in the regression.

Finally, the evidence about the effect of the level of political integration in the measured level of city size inequality is mixed. In the Urbanisation Model, systems of cities at the regional scale appear more uneven than systems of cities at the national scale, whereas in the Complete Model, systems of cities at the national scale are systematically more uneven than systems of cities at other scales.

## 4. CONCLUSION

The rank-size rule (or Zipf's law) for cities is a fascinating urban regularity and an unusual example in social research where so many different attempts have been made to explore, confirm or reject the same prediction with a variety of quantitative experiments. This is no wonder then that, to our knowledge, this is the only aspect of social theory (along with economies of agglomerations [Melo, 2009]) for which it is possible to perform a meta-analysis on a significant number of papers [Nitsch, 2005]. In the present paper, we introduced an innovative layer of cumulative knowledge and reproducibility by constructing a very large database of estimates, by making it freely available online, and by allowing the crowdsourcing of further estimates which we might have missed. Furthermore, we allow for the reproducibility and exploration of results by providing an interactive application along with the more standard format of a scientific paper. The scale of this literature review allows us to draw more robust conclusions because the coverage of estimates by continent, city definition, time periods and other categories is wider.

We conclude that a large part of the variations observed in the measure of Zipf's coefficient is unnecessary as it comes from the choice of different technical specifications in the way cities are defined. This leaves room for improvement in the harmonisation of data and the reporting of results. Improved results will then allow the confirmation or rejection of urbanisation theories and the prediction of future levels of size inequality. In the current form, the meta-analysis performed in this paper concludes that there is a persistent difference in the size distribution of areas of ancient and recent urbanisation, with the latter being more uneven. The structure of urban settlement thus proves rather independent from economic development and also very long to adjust to new transportation features. However, we have shown that the evolution of city size inequality over time and its relation to the total population was unclear or non-significant. This contradicts previous claims and restricts our confidence to predict future levels of urban unevenness with the current state of theory. In some cases however, for example shrinking countries, it would be very interesting to be able to forecast future levels of urban imbalance, because size inequalities are linked to broader socioeconomic and efficiency inequalities (because of urban scaling) which are important to the fair governance of societies.

Our conclusions thus call for a refinement of urban data, a greater level of precision in the reporting of results, including the specifications of regressions, a larger habit of sharing data to enhance the (re)production of cumulative knowledge, and finally a deeper investigation of theoretical explanations for the diversity of city size distributions in the world.

# 5. REFERENCES

Auerbach, F. (1913). Das Gesetz der Bevölkerungskonzentration. *Petermanns Geographische Mitteilungen*, *59*, 74-76.

Berry, B. J., & Garrison, W. L. (1958). Alternate explanations of urban rank-size relationships 1. *Annals of the Association of American Geographers*, *48*(1), 83-90.

Cottineau, C. (2014). *L'évolution des villes dans l'espace post-soviétique, Observation et modélisations*. Université Paris 1 – Panthéon-Sorbonne, Paris.

Cristelli, M., Batty, M., & Pietronero, L. (2012). There is more than a power law in Zipf. *Scientific reports*, *2*.

Eeckhout, J. (2004). Gibrat's law for (all) cities. *American Economic Review*, 1429-1451.

Fujita, M., Krugman P., Venables A. J. (1999), *The Spatial Economy: Cities, Regions and International Trade*, Cambridge, MA, MIT Press.

Gabaix, X., & Ibragimov, R. (2011). Rank– 1/2: a simple way to improve the OLS estimation of tail exponents. *Journal of Business & Economic Statistics*, *29*(1), 24-39.

González-Val, R. (2010). The Evolution of U.s. City Size Distribution from a Long-Term Perspective (1900–2000)*. *Journal of Regional Science*, *50*(5), 952–972.

Harris, C. D. (1970). *Cities of the Soviet Union: studies in their functions, size, density, and growth* (No. 5). Chicago: Association of American Geographers.

Jefferson, M. (1939). The law of the primate city. *Geographical review*, *29*, 226-232.

Johnson, G. A. (1980). Rank-size convexity and system integration: A view from archaeology. *Economic Geography*, 234-247.

Kumar, G., & Subbarayan, A. (2014). The temporal dynamics of regional city size distribution: Andhra Pradesh (1951-2001). *Journal of Mathematics and Statistics*, *10*(2), 221.

Luckstead, J., & Devadoss, S. (2014). A comparison of city size distributions for China and India from 1950 to 2010. *Economics Letters*, *124*(2), 290–295.

Melo, P. C., Graham, D. J., & Noland, R. B. (2009). A meta-analysis of estimates of urban agglomeration economies. *Regional science and urban Economics*, *39*(3), 332-342.

Moriconi-Ebrard, F. (1993). *L'urbanisation du monde: depuis 1950*. Economica.

Morrill, R. L. (1970). *The spatial organization of society* (pp. 175-184). Duxbury Press.

Nitsch, V. (2005). Zipf zipped. *Journal of Urban Economics*, *57*(1), 86-100.

Parr, J. B. (1985). A note on the size distribution of cities over time. *Journal of Urban Economics*, *18*(2), 199–212.

Pumain, D. (1997). Pour une théorie évolutive des villes. *Espace géographique*, *26*(2), 119-134.

Rosen, K. T., & Resnick, M. (1980). The size distribution of cities: an examination of the Pareto law and primacy. *Journal of Urban Economics*, *8*(2), 165–186.

Savage, S. H. (1997). Assessing departures from log-normality in the rank-size rule. *Journal of archaeological Science*, *24*(3), 233-244.

Soo, K. T. (2005). Zipf's Law for cities: a cross-country investigation. *Regional Science and Urban Economics*, *35*(3), 239–263.

Xu, Z., & Zhu, N. (2009). City size distribution in China: are large cities dominant? *Urban Studies*, *46*(10), 2159–2185.

Zipf, G. K. (1949). Human behavior and the principle of least effort.

# 6. APPENDIX

## Appendix A: List of studies used for the meta analysis

1. Alperovich, G. (1989). The distribution of city size: A sensitivity analysis. *Journal of Urban Economics*, *25*(1), 93–102.
2. Amalraj, V. C., & Subbarayan, A. (2014). Zipf's Law and Urban Dynamics in an Indian State: Kerala (1951-2001). *Journal of Applied Sciences*, *14*(24), 3604.
3. Anderson, G., & Ge, Y. (2005). The size distribution of Chinese cities. *Regional Science and Urban Economics*, *35*(6), 756–776. http://doi.org/10.1016/j.regsciurbeco.2005.01.003
4. Aragón, J. A. O., & Queiroz, V. dos S. (2014). The Zipf's law and the effects of free trade: The case of Guatemala. *EconomiA*, *15*(1), 82–99. http://doi.org/10.1016/j.econ.2014.03.007
5. Arribas-Bel, D., Gracia, F. S., & Ximénez-de-Embún, D. (2012). Kangaroos, cities and space: a first approach to the australian urban system. *Region et Developpement*, *36*, 165–187.
6. Batty, M. (2001). Polynucleated urban landscapes. *Urban Studies*, *38*(4), 635–655.
7. Benguigui, L., & Blumenfeld-Lieberthal, E. (2007). Beyond the power law – a new approach to analyze city size distributions. *Computers, Environment and Urban Systems*, *31*(6), 648–666. http://doi.org/10.1016/j.compenvurbsys.2006.11.002
8. Berry, B. J., & Okulicz-Kozaryn, A. (2012). The city size distribution debate: Resolution for US urban regions and megalopolitan areas. *Cities*, *29*, S17–S23.
9. Black, D., & Henderson, V. (2003). Urban evolution in the USA. *Journal of Economic Geography*, *3*(4), 343–372.
10. Bosker, M., Brakman, S., Garretsen, H., & Schramm, M. (2008). A century of shocks: The evolution of the German city size distribution 1925–1999. *Regional Science and Urban Economics*, *38*(4), 330–347. http://doi.org/10.1016/j.regsciurbeco.2008.04.002
11. Brakman, S., Garretsen, H., Van Marrewijk, C., & Van Den Berg, M. (1999). The return of Zipf: towards a further understanding of the rank-size distribution. *Journal of Regional Science*, *39*(1), 183–213.
12. Bretagnolle, A., Delisle, F., Mathian, H., & Vatin, G. (2015). Urbanization of the United States over two centuries: an approach based on a long-term database (1790–2010). *International Journal of Geographical Information Science*, (ahead-of-print), 1–18.
13. Bretagnolle, A., Giraud, T., & Mathian, H. (2008). La mesure de l'urbanisation aux Etats-Unis, des premiers comptoirs coloniaux aux Metropolitan Areas (1790-2000). *Cybergeo : European Journal of Geography*. http://doi.org/10.4000/cybergeo.19683
14. Bretagnolle, A., Mathian, H., Pumain, D., & Rozenblat, C. (2000). Long-term dynamics of European towns and cities : towards a spatial model of urban growth. *Cybergeo : European Journal of Geography*. http://doi.org/10.4000/cybergeo.566
15. Cameron, T. A. (1990). One-stage structural models to explain city size. *Journal of Urban Economics*, *27*(3), 294–307.
16. Catin, M., Cuenca, C., Kamal, A., & others. (2008). L'Évolution De La Structure Et De La Primatie Urbaines Au Maroc. *Région et Développement*, *27*, 215–223.
17. Cottineau, C. (2014). *L'évolution des villes dans l'espace post-soviétique, Observation et modélisations*. Université Paris 1 – Panthéon-Sorbonne, Paris. Retrieved from http://dx.doi.org/10.6084/m9.figshare.1348299
18. Crampton, G. (2005). The Rank-Size Rule in Europe-testing Zipf's law using European data.
19. Delgado, A. P., & Godinho, I. M. (2004). The evolution of city size distribution in Portugal: 1864-2001. Retrieved from http://www.econstor.eu/handle/10419/117123
20. Deliktas, E., Önder, A. Ö., & Karadag, M. (2013). The size distribution of cities and determinants of city growth in Turkey. *European Planning Studies*, *21*(2), 251–263.
21. Deurloo, M. C. (1972). De wet der urbane concentratie. *Tijdschrift Voor Economische En Sociale Geografie*, *63*(5), 306–314.
22. Dimou, M., & Schaffar, A. (2009). Urban hierarchies and city growth in the Balkans. *Urban Studies*. Retrieved from http://usj.sagepub.com/content/early/2009/09/04/0042098009344993.short
23. Dobkins, L. H., Ioannides, Y. M., & others. (2000). Dynamic evolution of the US city size distribution. *The Economics of Cities*, 217–260.
24. Eeckhout, J. (2004). Gibrat's Law for (All) Cities. *The American Economic Review*, *94*(5), 1429–1451.
25. Ettlinger, N., & Archer, J. C. (1987). City-size distributions and the world urban system in the twentieth century. *Environment and Planning A*, *19*(9), 1161–1174.
26. Ezzahid, E., & ElHamdani, O. (2015). Zipf's Law in the Case of Moroccan Cities. *Review of Urban & Regional Development Studies*, *27*(2), 118–133. http://doi.org/10.1111/rurd.12036
27. Fazio, G., & Modica, M. (2015). Pareto or Log-Normal? Best Fit and Truncation in the Distribution of All Cities*. *Journal of Regional Science*, *55*(5), 736–756. http://doi.org/10.1111/jors.12205
28. Gabaix, X. (1999). Zipf's law for cities: an explanation. *Quarterly Journal of Economics*, 739–767.

29. Gan, L., Li, D., & Song, S. (2006). Is the Zipf law spurious in explaining city-size distributions? *Economics Letters*, *92*(2), 256–262. http://doi.org/10.1016/j.econlet.2006.03.004
30. Gangopadhyay, K., & Basu, B. (2013). Evolution of Zipf's Law for Indian Urban Agglomerations Vis-à-Vis Chinese Urban Agglomerations. In F. Abergel, B. K. Chakrabarti, A. Chakraborti, & A. Ghosh (Eds.), *Econophysics of Systemic Risk and Network Dynamics* (pp. 119–129). Springer Milan. Retrieved from http://link.springer.com/chapter/10.1007/978-88-470-2553-0_8
31. Giesen, K., & Südekum, J. (2010). Zipf's law for cities in the regions and the country. *Journal of Economic Geography*, lbq019.
32. González-Val, R. (2010). The Evolution of U.s. City Size Distribution from a Long-Term Perspective (1900–2000)*. *Journal of Regional Science*, *50*(5), 952–972. http://doi.org/10.1111/j.1467-9787.2010.00685.x
33. Guérin-Pace, F. (1995). Rank-size distribution and the process of urban growth. *Urban Studies*, *32*(3), 551–562.
34. Gulden, T. R., & Hammond, R. A. (2012). Beyond Zipf: An agent-based understanding of city size distributions. In *Agent-based models of geographical systems* (pp. 677–704). Springer. Retrieved from http://link.springer.com/chapter/10.1007/978-90-481-8927-4_34
35. Holmes, T. J., & Lee, S. (2010). Cities as Six-by-Six-Mile Squares: Zipf's Law? *NBER*, 105–131.
36. Hongying, G., & Kangping, W. (2008). Zipf's law and influential factors of the Pareto exponent of the city size distribution: Evidence from China. *Frontiers of Economics in China*, *3*(1), 137–149.
37. Ignazzi, C.A., 2015, *Coevolution in the Brazilian system of cities*, thèse de Doctorat, Université Paris 1, Panthéon-Sorbonne, Paris, France.
38. Iyer, S. D. (2003). Increasing Unevenness in the Distribution of City Sizes in Post-Soviet Russia. *Eurasian Geography and Economics*, *44*(5), 348–367. http://doi.org/10.2747/1538-7216.44.5.348
39. Kamecke, U. (1990). Testing the rank size rule hypothesis with an efficient estimator. *Journal of Urban Economics*, *27*(2), 222–231.
40. Knudsen, T. (2001). Zipf's law for cities and beyond: The case of Denmark. *American Journal of Economics and Sociology*, *60*(1), 123–146.
41. Krakover, S. (1998). Testing the turning-point hypothesis in city-size distribution: the Israeli situation re-examined. *Urban Studies*, *35*(12), 2183–2196.
42. Krugman, P. (1996). Confronting the mystery of urban hierarchy. *Journal of the Japanese and International Economies*, *10*(4), 399–418.
43. Kumar, G., & Subbarayan, A. (2014). The temporal dynamics of regional city size distribution: Andhra Pradesh (1951-2001). *Journal of Mathematics and Statistics*, *10*(2), 221.
44. Le Gallo, J., & Chasco, C. (2008). Spatial analysis of urban growth in Spain, 1900–2001. *Empirical Economics*, *34*(1), 59–80.
45. Lepetit, B. (1990). Patterns of settlement and political changes: The French Revolution and the National Urban Hierarchy. In A. van der Woulde, A. Hayami, & J. de Vries, *Urbanization in History* (Clarendon Press). Oxford.
46. Lotka, A. J. (1925). *Elements of physical biology* (William & Wilkins). Baltimore. Retrieved from http://agris.fao.org/agris-search/search.do?recordID=US201300526822
47. Luckstead, J., & Devadoss, S. (2014). A comparison of city size distributions for China and India from 1950 to 2010. *Economics Letters*, *124*(2), 290–295. http://doi.org/10.1016/j.econlet.2014.06.002
48. Malecki, E. J. (1980). Growth and change in the analysis of rank-size distributions: empirical findings. *Environment and Planning A*, *12*(1), 41–52.
49. Mills, E. S., Becker, C. M., & Verma, S. (1986). *Studies in Indian urban development*. Oxford University Press. Retrieved from https://books.google.fr/books?hl=fr&lr=&id=btVZV3qvwFwC&oi=fnd&pg=PP227&dq=mills+becker+studies+in+indian+urban+development&ots=2swEsNh5Xk&sig=E-72ZjhD-EYjE6A0GZwVKbA8Lrs
50. Mirucki, J. (1986). Planned Economic Development and Loglinearity in the Rank-Size Distribution of Urban Systems: The Soviet Experience. *Urban Studies*, *23*(2), 151–156.
51. Modica, M., Reggiani, A., & Nijkamp, P. (2015). A Comparative Analysis of Gibrat's and Zipf's Law on Urban Population. http://doi.org/10.2139/ssrn.2611960
52. Moore, F. T. (1959). A note on city size distributions. *Economic Development and Cultural Change*, *7*(4), 465–466.
53. Moriconi-Ebrard, F. (1993). *L'urbanisation du monde: depuis 1950*. Economica.
54. Moro, S., & Santos, R. (2013). *The characteristics and evolution of the Brazilian spatial urban system: empirical evidences for the long-run, 1970-2010* (Textos para Discussão Cedeplar-UFMG No. 474). Cedeplar, Universidade Federal de Minas Gerais. Retrieved from http://econpapers.repec.org/paper/cdptexdis/td474.htm
55. Morudu, H., & Plessis, D. du. (2013). Economic and demographic trends of municipalities in South Africa: An application of Zipf's rule. *Town and Regional Planning*, *63*(1), 24–36.
56. Naudé, W. A., & Krugell, W. F. (2003). An Inquiry into Cities and their Role in Subnational Economic Growth in South Africa. *Journal of African Economies*, *12*(4), 476–499. http://doi.org/10.1093/jae/12.4.476

57. Nishiyama, Y., Osada, S., & Sato, Y. (2008). OLS estimation and the t test revisited in rank-size rule regression. *Journal of Regional Science*, *48*(4), 691–716.
58. Okabe, A. (1979). An expected rank-size rule: A theoretical relationship between the rank size rule and city size distributions. *Regional Science and Urban Economics*, *9*(1), 21–40.
59. Parr, J. B. (1985). A note on the size distribution of cities over time. *Journal of Urban Economics*, *18*(2), 199–212.
60. Parr, J. B., & Jones, C. (1983). City size distributions and urban density functions: some interrelationships. *Journal of Regional Science*, *23*(3), 283–307.
61. Paulus, F. (2004). *Coévolution dans les systèmes de villes: croissance et spécialisation des aires urbaines françaises de 1950 à 2000*. Université Panthéon-Sorbonne-Paris I. Retrieved from https://tel.archives-ouvertes.fr/tel-00008053/
62. Popov, V. R. (1974). Investigation of the System of Urban Places of Crimea Oblast. *Soviet Geography*, *15*(1), 19–23.
63. Pumain, D., Swerts, E., Cottineau, C., Vacchiani-Marcuzzo, C., Ignazzi, A., Bretagnolle, A., … Baffi, S. (2015). Multilevel comparison of large urban systems. *Cybergeo : European Journal of Geography*. http://doi.org/10.4000/cybergeo.26730
64. Rosen, K. T., & Resnick, M. (1980). The size distribution of cities: an examination of the Pareto law and primacy. *Journal of Urban Economics*, *8*(2), 165–186.
65. Rozenfeld, H. D., Rybski, D., Gabaix, X., & Makse, H. A. (2011). The Area and Population of Cities: New Insights from a Different Perspective on Cities. *American Economic Review*, *101*(5), 2205–2225. http://doi.org/10.1257/aer.101.5.2205
66. Schaffar, A., & Dimou, M. (2012). Rank-size city dynamics in China and India, 1981–2004. *Regional Studies*, *46*(6), 707–721.
67. Schaffar, A., & Nassori, D. (2016). La croissance urbaine marocaine : convergence vs concentration. *Revue économique*, *67*(2), 207–226.
68. Shepotylo, O. (2012). Cities in transition. *Comparative Economic Studies*, *54*(3), 661–688.
69. Singer, H. W. (1936). The' courbe des populations.' A parallel to Pareto's Law. *The Economic Journal*, 254–263.
70. Song, S., & Zhang, K. H. (2002). Urbanisation and city size distribution in China. *Urban Studies*, *39*(12), 2317–2327.
71. Soo, K. T. (2005). Zipf's Law for cities: a cross-country investigation. *Regional Science and Urban Economics*, *35*(3), 239–263.
72. Soo, K. T. (2007). Zipf's Law and urban growth in Malaysia. *Urban Studies*, *44*(1), 1–14.
73. Suarez-Villa, L. (1980). Rank size distribution, city size hierarchies and the Beckmann model: some empirical results. *Journal of Regional Science*, *20*(1), 91–95.
74. Subbarayan, A., Kumar, G., & Amalraj, V. C. (2011). The temporal and spatial dynamics of regional city-size distribution: Tamilnadu (1951-2001). *Int. J. Agric. Stat. Sci*, *7*, 535–554.
75. Swerts, E. (2013). *Les Systèmes de Villes en Inde et en Chine.* Paris 1 Pantheon-Sorbonne. Retrieved from http://www.theses.fr/s90230
76. Vacchiani-Marcuzzo, C. (2005). *Mondialisation et système de villes: les entreprises étrangères et l'évolution des agglomérations sud-africaines*. Université Panthéon-Sorbonne-Paris I. Retrieved from https://tel.archives-ouvertes.fr/tel-00011351/
77. Veneri, P., & others. (2013). *On city size distribution: evidence from OECD functional urban areas*. Oecd Publishing. Retrieved from http://ideas.repec.org/p/oec/govaab/2013-27-en.html
78. Ward, B. (1963). City structure and interdependence. In *Papers of the Regional Science Association* (Vol. 10, pp. 207–221). Springer. Retrieved from http://link.springer.com/article/10.1007/BF01934688
79. Xu, Z., & Zhu, N. (2009). City size distribution in China: are large cities dominant? *Urban Studies*, *46*(10), 2159–2185.
80. Your. (2003). Alaniz izmenenii v yerarhii gorodov Rossii s ispolzovanyem pravila rang-razmer. In D. Eckert & V. Kolossov, *Krupnye goroda y vyzovy urbanizaciya* (Institut Geog. RAN, pp. 72–80). Moscow.
81. Ziqin, W. (2016). Zipf Law Analysis of Urban Scale in China. *Asian Journal of Social Science Studies*, *1*(1), 53.

## Appendix B: 72 estimations confirming Zipf's law (0.99 < α < 1.01)

| Alpha | Territory | Date | Urban Age | N | City Definition | Pop min | Discipline | R2 | Total Pop (x1000) | Reference |
|---|---|---|---|---|---|---|---|---|---|---|
| 1.009 | Sichuan | 2000 | OLD | | LocalUnit | | ECO | | 118811.73 | Hongying Kangping 2008 |
| 1.009 | United States of America | 2010 | RECENT | | MetroArea | 500000 | SOC | 96.2 | 309876.17 | Berry Okulicz-Kozaryn 2012 |

| Alpha | Territory | Date | Urban Age | N | City Definition | Pop min | Discipline | R2 | Total Pop (x1000) | Reference |
|---|---|---|---|---|---|---|---|---|---|---|
| 1.008 | France | 2000 | OLD | 109 | MorphoCity | 51000 | | 99 | 59387.18 | Graham 2005 |
| 1.007 | Iran | 1966 | OLD | 50 | LocalUnit | sample | ECO & SOC | | 25624.65 | Rosen Resnick 1980 |
| 1.005 | United States of America | 1940 | RECENT | 160 | MetroArea | 77886 | | 98.2 | | Dobkins Ioannides 2000 |
| 1.005 | American MidWest | 1940 | RECENT | 566 | LocalUnit | 2500 | SOC & PHYS | 99.4 | | Malecki 1980 |
| 1.005 | United States of America | 2000 | RECENT | 250 | MorphoCity | 150000 | | | 282895.74 | Gulden Hammond 2012 |
| 1.004 | Egypt | 1996 | OLD | 127 | LocalUnit | 0 | ECO & SOC | | 63595.63 | Soo 2005 |
| 1.004 | Shandong | 2000 | OLD | | LocalUnit | | ECO | | 118811.73 | Hongying Kangping 2008 |
| 1.004 | Shandong | 2000 | OLD | | LocalUnit | | | | 118811.73 | Ziqin 2016 |
| 1.004 | Zhejiang | 2005 | OLD | | LocalUnit | | | | 1208919.51 | Ziqin 2016 |
| 1.003 | South Africa | 1970 | RECENT | 50 | LocalUnit | sample size | ECO & SOC | | 22502.5 | Rosen Resnick 1980 |
| 1.003 | Sichuan | 1997 | OLD | | LocalUnit | | ECO | | 1177796.64 | Hongying Kangping 2008 |
| 1.003 | Spain | 2001 | OLD | 76 | MetroArea | 50000 | ECO, SOC & PHYS | 98 | 41230.52 | Veneri 2013 |
| 1.002 | Egypt | 1947 | OLD | | LocalUnit | 20000 | ECO & SOC | | | Parr 1985 |
| 1.002 | United States of America | 1970 | RECENT | 300 | MetroArea | sample size | ECO & SOC | | 209485.81 | Alperovich 1989 |
| 1.002 | United States of America | 1980 | RECENT | 270 | MetroArea | sample size | ECO & SOC | | 229588.21 | Alperovich 1989 |
| 1.002 | Andhra Pradesh | 2001 | OLD | 188 | LocalUnit | 10000 | | 93 | 1192558.3 | Kumar Subbarayan 2014 |
| 1.001 | United States of America | 1950 | RECENT | 162 | MetroArea | 101013 | | 97.8 | 157813.04 | Dobkins Ioannides 2000 |
| 1.001 | Austria | 1981 | OLD | | LocalUnit | 5000 | ECO & SOC | | 7591.62 | Parr 1985 |
| 1.001 | United States of America | 2000 | RECENT | 135 | MetroArea | 280843 | ECO | 98.5 | 282895.74 | Eeckhout 2004 |
| 1.001 | Mexico | 2001 | RECENT | 77 | MetroArea | 50000 | ECO, SOC & PHYS | 95 | 104239.56 | Veneri 2013 |
| 1 | France | 1931 | OLD | 502 | MorphoCity | 5000 | SOC & PHYS | 99.5 | | Guerin-Pace 1995 |
| 1 | France | 1931 | OLD | 1115 | MorphoCity | 2000 | SOC & PHYS | 99.7 | | Guerin-Pace 1995 |
| 1 | Mexico | 1950 | RECENT | | MorphoCity | 10000 | | 98.4 | 28012.56 | Moriconi-Ebrard 1993 |
| 1 | United Kingdom | 1956 | OLD | 25 | MetroArea | sample size | | 90 | 51315.72 | Ward 1962 |
| 1 | Vietnam | 1960 | OLD | | MorphoCity | 10000 | | 95.2 | 32670.62 | Moriconi-Ebrard 1993 |
| 1 | Azerbaijan | 1970 | OLD | | MorphoCity | 10000 | | 86.6 | 5178.16 | Moriconi-Ebrard 1993 |
| 1 | Bangladesh | 1970 | OLD | | MorphoCity | 10000 | | 96.7 | 65048.7 | Moriconi-Ebrard 1993 |
| 1 | United States of America | 1970 | RECENT | 50 | MetroArea | sample size | ECO & SOC | | 209485.81 | Rosen Resnick 1980 |
| 1 | Denmark | 1980 | OLD | | MorphoCity | 10000 | | 94.8 | 5123.44 | Moriconi-Ebrard 1993 |
| 1 | Netherlands | 1980 | OLD | | MorphoCity | 10000 | | 99.4 | 14103.28 | Moriconi-Ebrard 1993 |
| 1 | Sudan | 1980 | RECENT | | MorphoCity | 10000 | | 97.3 | 4701.36 | Moriconi-Ebrard 1993 |
| 1 | Vietnam | 1980 | OLD | | MorphoCity | 10000 | | 98.7 | 54372.52 | Moriconi-Ebrard 1993 |
| 1 | Indonesia | 1990 | OLD | 193 | MorphoCity | 0 | ECO & SOC | | 181436.82 | Soo 2005 |
| 0.999 | Andhra Pradesh | 1981 | OLD | 156 | LocalUnit | 5000 | | 93.8 | 1088334.84 | Kumar Subbarayan 2014 |
| 0.999 | Morocco | 2010 | RECENT | 117 | MorphoCity | | ECO | 98 | 32107.74 | Schaffar Nassori 2016 |
| 0.997 | Japan | 2000 | OLD | 113 | MetroArea | 10000 | SOC & PHYS | 95.2 | 125714.67 | Nishiyama 2008 |

| Alpha | Territory | Date | Urban Age | N | City Definition | Pop min | Discipline | R2 | Total Pop (x1000) | Reference |
|---|---|---|---|---|---|---|---|---|---|---|
| 0.996 | United States of America | 1991 | RECENT | 130 | MetroArea | 100000 | ECO & SOC | | 255367.16 | Krugman 1996 |
| 0.996 | Russia | 2010 | OLD | 260 | MorphoCity | 50000 | | 96.42 | 143158.1 | Cottineau 2014 |
| 0.995 | United States of America | 1991 | RECENT | 135 | MetroArea | 250000 | ECO | 98.6 | 255367.16 | Gabaix 1999 |
| 0.995 | OECD29 | 2001 | | | MetroArea | 50000 | ECO, SOC & PHYS | 97 | 1192558.3 | Veneri 2013 |
| 0.994 | United States of America | 1980 | RECENT | 121 | MetroArea | 100000 | ECO & SOC | 91 | 229588.21 | Cameron 1990 |
| 0.994 | Portugal | 1991 | OLD | 110 | LocalUnit | 2000 | | 94.7 | 9909.57 | Delgado Godinho 2004 |
| 0.994 | United States of America | 2000 | RECENT | | MetroArea | 500000 | SOC | | 282895.74 | Berry Okulicz-Kozaryn 2012 |
| 0.993 | Portugal | 1981 | OLD | 110 | LocalUnit | 2000 | | 93.1 | 9824.24 | Delgado Godinho 2004 |
| 0.992 | USSR | 1979 | OLD | 1116 | LocalUnit | 20000 | ECO | 98.4 | | Shepotylo 2012 |
| 0.992 | Luxembourg | 1980 | OLD | 116 | LocalUnit | 0 | | | 364.04 | Modica al 2015 |
| 0.992 | Poland | 2001 | OLD | 58 | MetroArea | 50000 | ECO, SOC & PHYS | 99 | 38466.54 | Veneri 2013 |
| 0.992 | Czech Republic | 2007 | OLD | 21 | LocalUnit | 50000 | ECO | 93 | 10330.49 | Shepotylo 2012 |
| 0.991 | Portugal | 1991 | OLD | 122 | LocalUnit | 2000 | | 96.1 | 9909.57 | Delgado Godinho 2004 |
| 0.99 | United States of America | 1900 | RECENT | 64 | MetroArea | sample size | ECO & SOC | 97.5 | | Black Henderson 2003 |
| 0.99 | United States of America | 1910 | RECENT | 510 | MetroArea | 10000 | SOC | 99 | | Bretagnolle al 2008 |
| 0.99 | United States of America | 1920 | RECENT | 149 | MetroArea | 74161 | | 99 | | Dobkins Ioannides 2000 |
| 0.99 | France | 1950 | OLD | | MorphoCity | 10000 | | 98.6 | 41879.61 | Moriconi-Ebrard 1993 |
| 0.99 | Kazakhstan | 1950 | RECENT | | MorphoCity | 10000 | | 95.7 | 6703 | Moriconi-Ebrard 1993 |
| 0.99 | Peru | 1950 | RECENT | | MorphoCity | 10000 | | 91.7 | 7727.74 | Moriconi-Ebrard 1993 |
| 0.99 | Bangladesh | 1960 | OLD | | MorphoCity | 10000 | | 95 | 48200.7 | Moriconi-Ebrard 1993 |
| 0.99 | Brazil | 1970 | RECENT | | MorphoCity | 10000 | | 99.4 | 95982.45 | Moriconi-Ebrard 1993 |
| 0.99 | Ghana | 1970 | RECENT | | MorphoCity | 10000 | | 94.6 | 8596.98 | Moriconi-Ebrard 1993 |
| 0.99 | Greece | 1970 | OLD | | MorphoCity | 10000 | | 90.6 | 8778.68 | Moriconi-Ebrard 1993 |
| 0.99 | Japan | 1970 | OLD | | MorphoCity | 10000 | | 99.3 | 103707.54 | Moriconi-Ebrard 1993 |
| 0.99 | Sudan | 1970 | RECENT | | MorphoCity | 10000 | | 97 | 3647.1 | Moriconi-Ebrard 1993 |
| 0.99 | Mexico | 1980 | RECENT | 54 | LocalUnit | 100000 | ECO & SOC | | 69330.97 | Kamecke 1990 |
| 0.99 | Azerbaijan | 1980 | OLD | | MorphoCity | 10000 | | 88.6 | 6163.99 | Moriconi-Ebrard 1993 |
| 0.99 | Finland | 1980 | OLD | | MorphoCity | 10000 | | 99.1 | 4779.45 | Moriconi-Ebrard 1993 |
| 0.99 | Indonesia | 1980 | OLD | | MorphoCity | 10000 | | 98.4 | 147490.37 | Moriconi-Ebrard 1993 |
| 0.99 | Philippines | 1980 | OLD | | MorphoCity | 10000 | | 97.2 | 47396.97 | Moriconi-Ebrard 1993 |
| 0.99 | Poland | 1980 | OLD | | MorphoCity | 10000 | | 99.7 | 35782.86 | Moriconi-Ebrard 1993 |
| 0.99 | Romania | 1980 | OLD | | MorphoCity | 10000 | | 97.9 | 22612.15 | Moriconi-Ebrard 1993 |
| 0.99 | Tanzania | 1988 | RECENT | 32 | LocalUnit | 0 | ECO & SOC | | 23914.85 | Soo 2005 |
| 0.99 | Germany | 2000 | OLD | 158 | MorphoCity | 50000 | | 99.2 | 81895.93 | Graham 2005 |

# Appendix C: Visual Appearance of the Online application MetaZipf

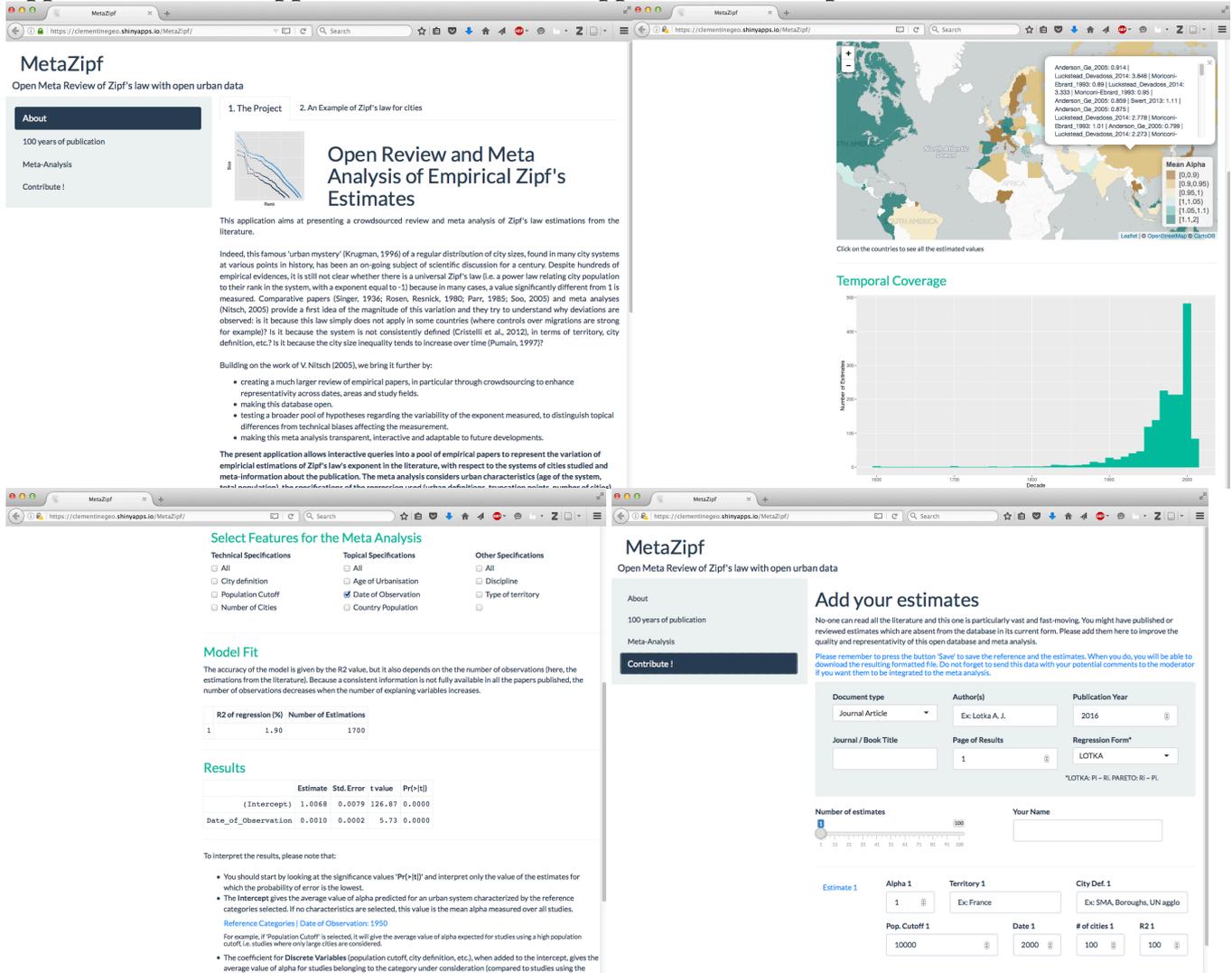

Source: https://clementinegeo.shinyapps.io/MetaZipf